\documentclass[copyright,creativecommons]{eptcs}
\providecommand{\event}{FTSCS 2012} 

\usepackage{breakurl}             

\usepackage{amsmath,amssymb,amsfonts}
\usepackage{latexsym,mathrsfs,stmaryrd,mathtools}
\usepackage{alltt,paralist}
\usepackage{relsize} 
\usepackage{hyperref}
\usepackage{wrapfig}
\usepackage{graphicx}

\usepackage{shortvrb}  
\usepackage{dsfont}    

\newcommand{\MA}{\mathcal{M}\!\!\mathcal{A}}
\newcommand{\attone}{\mbox{$att_1$}}
\newcommand{\attn}{\mbox{$att_n$}}
\newcommand{\sn}{\mbox{$s_n$}}

\newcommand{\sonex}{\mbox{$s_1$}}

\title{PALS-Based Analysis of an Airplane\\ Multirate Control System
  in Real-Time Maude} 

\author{Kyungmin Bae \quad\quad Joshua Krisiloff \quad\quad Jos\'e Meseguer
\institute{Department of Computer Science \\ University of Illinois at
  Urbana-Champaign} 
\and
Peter Csaba {\"O}lveczky
\institute{Department of Informatics \\  University of Oslo}
}
\def\titlerunning{Formal Specification and Analysis of an Airplane Turning Control System}
\def\authorrunning{K. Bae, J. Krisiloff, J. Meseguer, and P. C. {\"O}lveczky}
\begin{document}
\maketitle

\MakeShortVerb{\@}    

\begin{abstract}
Distributed cyber-physical systems (DCPS) are pervasive in areas such as aeronautics
and ground transportation systems, including the case of distributed hybrid systems.
 DCPS design and verification is quite challenging because of asynchronous communication, network delays, and clock skews. Furthermore, their model checking verification typically becomes unfeasible due to the huge state space explosion caused by the system's concurrency.
 The PALS  (``physically asynchronous, logically synchronous'') methodology has been proposed to reduce the design and verification of a DCPS 
 to the much simpler task of designing and verifying its underlying synchronous version.
The original PALS methodology assumes a single logical period,
but Multirate PALS extends it to deal with multirate DCPS in which components may operate with different logical periods.
This paper shows how Multirate PALS can be applied to formally verify a nontrivial multirate DCPS. We use Real-Time Maude to formally specify a multirate distributed hybrid system consisting of an airplane maneuvered by a pilot who turns the airplane according to a specified angle through a distributed control system.
Our formal analysis revealed that the original design was ineffective in achieving a smooth turning maneuver,
and led to a redesign of the system that satisfies the desired correctness properties.
This shows that the Multirate PALS methodology is not only effective for formal DCPS verification, but can also be used effectively in the DCPS \emph{design} process, even before properties~are~verified.  
\end{abstract}

\section{Introduction}

Distributed cyber-physical systems (DCPS) are pervasive in areas such as aeronautics, ground transportation systems, medical systems, and so on; they include, in particular, the case of distributed hybrid systems, whose continuous dynamics is governed by differential equations.  DCPS design and verification is quite challenging, because to the usual complexity of a non-distributed CPS one has to add the additional complexities of asynchronous communication, network delays, and clock skews, which can easily lead a DCPS into inconsistent states.  In particular, any hopes of applying model checking
verification techniques in a direct manner to a DCPS look rather dim, due to the typically huge state space explosion
caused by the system's concurrency.

For these reasons, we and other colleagues at UIUC and Rockwell-Collins Corporation have been developing the 
\emph{physically asynchronous but logically synchronous}
PALS methodology \cite{icfem10,dasc09}, which can drastically reduce the system complexity of a DCPS so as to make it
amenable to model checking verification (for a comparison of PALS with related methodologies see \cite{steiner2011tta}). The PALS methodology applies to the frequently occurring case of
a DCPS whose implementation must be asynchronous due to physical constraints and for fault-tolerance reasons,
but whose \emph{logical} design requires that the system components should act together  in a virtually synchronous way.
For example, distributed control systems are typically of this nature.
The key idea of PALS is to reduce the design
and verification of a DCPS of this kind to the much simpler task%
\footnote{
For a simple avionics case study 
in~\cite{pals-tcs}, the
number of system states  for their 
simplest possible distributed version with perfect clocks and no
network delays was 3,047,832,  but the PALS
pattern reduced the number of states 
to a mere 185.}
 of designing and verifying the
idealized synchronous system that should be realized in a distributed and asynchronous way.
  This is achieved
by a \emph{model transformation}  
$\mathcal{E} \;\mapsto\; \mathcal{A}(\mathcal{E},\Gamma)$
 that maps a
synchronous design  $\mathcal{E}$ to a distributed implementation  $\mathcal{A}(\mathcal{E},\Gamma)$ which is
\emph{correct-by construction} and that, as shown in \cite{icfem10,pals-tcs}, is \emph{bisimilar} to the
synchronous system $\mathcal{E}$.   This bisimilarity is the essential feature allowing the desired, drastic reduction
in system complexity and making model checking verification feasible: since bisimilar systems satisfy the exact same temporal
logic properties, we can verify that the asynchronous system $\mathcal{A}(\mathcal{E},\Gamma)$ satisfies a temporal logic
property $\varphi$ (which typically would be impossible to model check directly on  $\mathcal{A}(\mathcal{E},\Gamma)$)
by verifying the same property $\varphi$ on the vastly simpler synchronous system~$\mathcal{E}$.

The original PALS methodology presented in \cite{icfem10,dasc09}  assumes 
a \emph{single logical period}, during  which
all components of the DCPS must communicate with each other and transition to their next states.   However,
a DCPS such as a distributed control system may have components that, for physical reasons, must operate with
different periods, even though those periods may all divide an overall longer period.  That is, many such
systems, although still having to be virtually synchronous for their correct behavior, are in fact \emph{multirate}
systems, with some components running at a faster rate than others.  An interesting challenge is how to extend 
PALS 
 to multirate DCPS.  This challenge has been given two different answers.  On the one hand,
an engineering solution for  Multirate PALS based on the AADL modeling  language has been proposed by
Al-Nayeem et al.\ in \cite{abdullah-multirate}.  On the other hand, three of us have
defined in \cite{bae-facs12}
a mathematical model of a multirate synchronous system $\mathcal{E}$, and have formally defined a
model transformation 
\[\mathcal{E} \;\mapsto\; \MA(\mathcal{E},T,\Gamma)\]
that generalizes to multirate systems the original single-rate PALS transformation defined in \cite{icfem10,pals-tcs}.
As before, we have proved in \cite{bae-facs12}
that  $ \MA(\mathcal{E},T,\Gamma)$ is a correct-by-construction implementation of  $\mathcal{E}$,
and that $\mathcal{E}$ and $\MA(\mathcal{E},T,\Gamma)$ are \emph{bisimilar}, making it possible
to verify temporal logic properties about $\MA(\mathcal{E},T,\Gamma)$ on the much simpler system
 $\mathcal{E}$.

 But how \emph{effective} is Multirate PALS in practice?  Can it be applied to formally verify important properties of a nontrivial multirate CPS such as a distributed hybrid system?  The main goal of this paper is to show that the answer is an emphatic \emph{yes}. We use Real-Time Maude \cite{journ-rtm} to formally specify in detail a multirate distributed hybrid system consisting of an airplane maneuvered by a pilot, who turns the airplane according to a specified angle $\alpha$ through a distributed control system with effectors located in the airplane's wings and rudder.  Our formal analysis revealed that the original design had control laws that were ineffective in achieving a smooth turning maneuver.  This led to a redesign of the system with new control laws which, as verified in Real-Time Maude by model checking, satisfies the desired correctness properties.  This shows that the Multirate PALS methodology is not only effective for formal DCPS verification, but, when used together with a tool like Real-Time Maude, can also be used effectively in the DCPS \emph{design} process, even before properties are verified.  To the best of our knowledge, this is the first time that the Multirate PALS methodology  has been applied to the model checking verification of a  DCPS.  In this sense, this paper complements our companion paper \cite{bae-facs12}, where the mathematical foundations of Multirate PALS were developed in detail, but where only a brief summary of some of the results presented here was given.
 
 This paper is organized as follows.
Section~\ref{sec:prelim}  explains  Multirate PALS and Real-Time Maude.
Section~\ref{sec:airplane-control} describes a simple model of an airplane turning control system 
whose continuous dynamics is governed by differential equations.
Section~\ref{sec:framework} presents a modeling framework for multirate ensembles in Real-Time Maude,
and Section~\ref{sec:airplane-model} then formally specifies the airplane turning control system using 
the ensemble  framework.
Section~\ref{sec:analysis} illustrates Real-Time Maude-based verification of the airplane turning control system.
Finally, Section~\ref{sec:concl} gives some concluding remarks.

\section{Preliminaries on  Multirate PALS and Real-Time Maude}
\label{sec:prelim}

\subsection{Multirate PALS}
In many distributed real-time systems, such as automotive and avionics systems, the
system design is essentially  
a \emph{synchronous design} that must be realized in a distributed setting. 
Both design and verification of such \emph{virtually synchronous}
distributed real-time systems is very hard 
because of asynchronous communication, network delays, clock skews,
and because the state space explosion caused by the system's
concurrency can make it unfeasible to apply model checking to verify
required properties. 
The (single-rate) PALS (``physically asynchronous, logically
synchronous'') formal design pattern~\cite{icfem10,dasc09}  reduces
the design and verification of  
  a  virtually synchronous distributed real-time system to the much
  simpler task of designing and verifying its  
synchronous version,  provided that the network infrastructure can
guarantee bounds on the messaging delays and the skews of the local
clocks.

%
We have 
recently developed Multirate PALS \cite{bae-facs12},
which extends PALS
to  hierarchical \emph{multirate} systems
in which  controllers with the same rate  communicate with
each other and with a number of faster components. 
As is common for hierarchical control
systems~\cite{abdullah-multirate}, 
we assume that the period of the higher-level controllers is a multiple of the period of each fast component.
%
Given  a multirate synchronous design $\mathfrak{E}$,  bounds
$\Gamma$ on the network transmission times and clock skews, and
function 
$T$ assigning to each distributed component its period, Multirate
PALS
defines a transformation $(\mathfrak{E}, T, \Gamma) \mapsto
\MA(\mathfrak{E}, T, \Gamma)$
mapping each synchronous design $\mathfrak{E}$ 
to a specification $\MA(\mathfrak{E}, T, \Gamma)$ of the
corresponding distributed multirate real-time system. 
In~\cite{bae-facs12} we formalize Multirate PALS
and show that the  synchronous design $\mathfrak{E}$ and the
asynchronous distributed model $\MA(\mathfrak{E}, T, \Gamma)$ satisfy
the same temporal logic properties.

\paragraph{Multirate Synchronous Models.}

In Multirate PALS, the synchronous design is formalized as the synchronous composition of a collection of \emph{typed machines},  
 an  \emph{environment}, and a \emph{wiring diagram} that  connects the machines. 
A \emph{typed machine}  $M$ is a tuple 
$M = (\mathcal{D}_i, S, \mathcal{D}_o,\delta_{M})$,
 where $\mathcal{D}_i=D_{i_1}\times \cdots \times D_{i_n}$ is an \emph{input set},  
 $S$ is  a  set of \emph{states},
$\mathcal{D}_o=D_{o_1}\times \cdots \times D_{o_m}$ is an \emph{output set},  
and $\delta_{M}\subseteq (D_i\times S)  \times (S\times D_o)$ is 
a  \emph{transition relation}.  
Such a machine $M$ has $n$ input ports and $m$ output ports.

To compose a collection of machines with different rates into a
synchronous system in which all components perform one transition in
lock-step in each iteration of the system, we  ``slow down'' the
faster components so that all components run at the slow rate in the
synchronous composition.  A fast machine that is slowed, or
\emph{decelerated}, by a factor $k$ \emph{performs $k$ internal
transitions} in  one synchronous step. Since the fast machine consumes 
an input and produces an output in each of these internal
steps, the decelerated machine consumes (resp.\ produces) $k$-tuples
of inputs (resp.\ outputs) in each synchronous  step.
A $k$-tuple output
from the fast machine must therefore be \emph{adapted} so that it can be read by
the slow component. That is, the $k$-tuple  must be transformed to a
single value (e.g., the average of the $k$ values, the last 
 value, or any other function of the $k$ values); this
transformation is formalized as an \emph{input adaptor}. Likewise,  the single output from a slow component
must be transformed to a $k$-tuple of inputs to the fast machine;
this  is also done by input adaptors which 
may, for example,   transform an input   $d$ to a $k$-tuple $(d, \bot,
\ldots, \bot)$ for some ``don't care'' value $\bot$. 
Formally, an input adaptor $\alpha_M$ for a typed machine $M = (\mathcal{D}_i, S, \mathcal{D}_o,\delta_{M})$
 is a family of functions 
 $\alpha_M = \{ \alpha_k : D_k' \to D_{i_k}\}_{k \in \{1,\ldots,n\}}$,
each of which determines a desired value from an output $D_k'$ of another typed machine.

\begin{figure}[htb]
\centering
\includegraphics[trim=0.4cm 0.4cm 0.35cm 0.4cm, clip=true,width=0.6\textwidth]{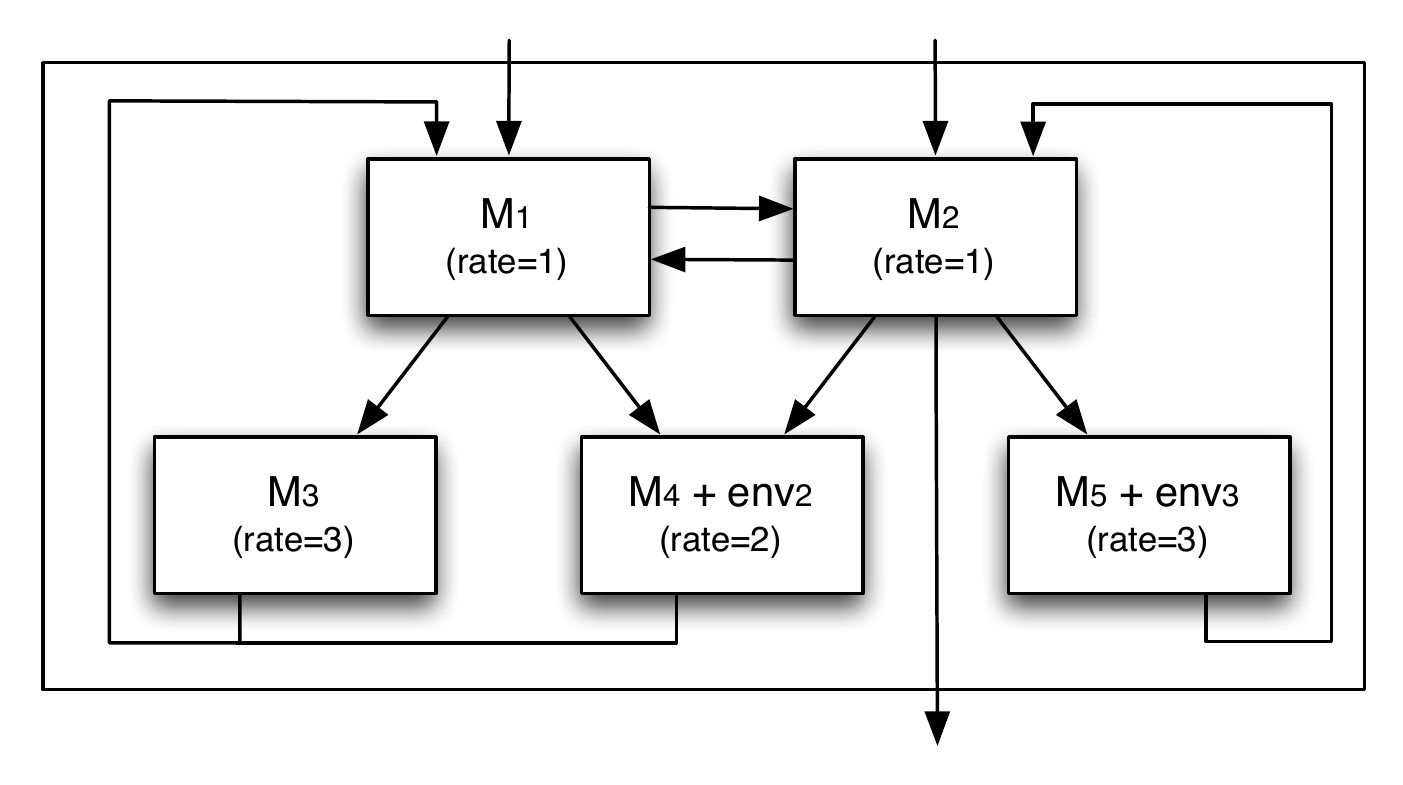}    
\caption{A multirate machine ensemble.
$M_1$ and $M_2$ are controller machines,
and $\mathit{env}_2$ and $\mathit{env}_3$ are local environments 
with faster rates, hidden from high-level controllers.
}  \label{fig:pals2}
\end{figure}

Typed machines (with rate assignments and input
adaptors)  can be ``wired together'' into \emph{multirate machine ensembles} as shown in 
 Figure~\ref{fig:pals2},
 where ``local'' fast environments are  integrated with their corresponding fast machines.
 A multirate machine ensemble is a tuple 
\[
\mathfrak{E} = (J_S \cup J_F \cup \{e\}, \{M_j\}_{j\in J_s \cup J_F}, E, \mathit{src}, \mathit{rate}, \mathit{adaptor})
\]
 where:
\begin{inparaenum}[(i)]
	\item $J_S$ (resp., $J_F$) is a  finite set of \emph{indices} for controller components (resp., fast components),
	and $e\not\in J_S \cup J_F$ is the \emph{environment index},
 	\item $\{M_j\}_{j\in J_S \cup J_F}$ is  a family of typed machines,
	\item the \emph{environment} is a  pair  $E=(\mathcal{D}^e_i, \mathcal{D}^e_o)$, 
	with $\mathcal{D}^{e}_i$   the environment's \emph{input set} and $\mathcal{D}^{e}_o$  its \emph{output set},
	\item $src$ is  a function that  assigns to each input port $(j, n)$ (input port $n$ of machine $j$) its ``source,''
	such that  there are no connections between fast machines,
	\item $\mathit{rate}$ is a function assigning to each fast machine a value denoting 
	how many times faster the machine runs compared to the controller machines, and
	\item $\mathit{adaptor}$ is a function that assigns an input adaptor to each $l \in J_F \cup J_S$.
\end{inparaenum}

The transitions of all  machines in an ensemble are performed simultaneously,
where each fast machine with rate $k$ performs $k$ ``internal
transitions''  in one  synchronous transition step.
If  a machine has a feedback wire to itself and/or to another
machine, then the  output becomes an input at the \emph{next} instant.  
The \emph{synchronous composition} of  a multirate ensemble  $\mathfrak{E}$ is therefore
equivalent to a \emph{single machine} $M_{\mathfrak{E}}$,  where
each state of $M_{\mathfrak{E}}$ consists of the states of its
subcomponents and of the contents in the feedback outputs,
as
explained in~\cite{bae-facs12,pals-tcs}.
The synchronous composition of the ensemble in Figure~\ref{fig:pals2}
is the  machine given by the outer box. 
Since  $M_{\mathfrak{E}}$ is itself a typed machine
which can appear as a component in another multirate ensemble, we can 
easily define hierarchical multirate systems.

\subsection{Real-Time Maude}
Real-Time Maude~\cite{journ-rtm} extends the Maude language and
tool~\cite{maude-book} to support the formal modeling analysis of
real-time systems in rewriting logic. In Real-Time Maude, the data
types of the system are defined by  a \emph{membership equational
logic}~\cite{maude-book} 
theory $(\Sigma, E)$  with $\Sigma$ a signature%
\footnote{That is, $\Sigma$ is a set  of declarations of \emph{sorts}, \emph{subsorts}, and
  \emph{function symbols}.} 
and $E$ a set of {\em confluent and terminating
conditional equations}; the system's  \emph{instantaneous}
(i.e., zero-time) local transitions are defined by (possibly
conditional) \emph{labeled instantaneous rewrite rules}%
\footnote{$E = E'\cup A$, where $A$ is a set of axioms such as associativity and commutativity,  
so that deduction is performed \emph{modulo} $A$. 
};
 and time elapse is modeled explicitly by \emph{tick rewrite rules} of the form 
\texttt{ crl\! [\(l\)]\! :\! \texttt{\char123}\(u\)\texttt{\char125}
  => \texttt{\char123}\(v\)\texttt{\char125} in time \(\tau\) if
  \(cond\)}, which  specifies a transition with duration $\tau$
from an instance of the term $\texttt{\char123}u\texttt{\char125}$ to the corresponding instance of the
term $\texttt{\char123}v\texttt{\char125}$.

The Real-Time Maude syntax  is fairly intuitive.
 A function symbol $f$  is declared with the syntax \texttt{op }$f$ @:@ $s_1$ \ldots $s_n$ @->@ $s$, 
 where $s_1\:\ldots\:s_n$ are the sorts of  its arguments,  and $s$  is its (value) \emph{sort}. 
Equations are written with syntax 
 \quad@eq@ $u$ @=@ $v$,\quad or \quad@ceq@ $u$ @=@ $v$ \,@if@\, \emph{cond}\quad
for conditional equations. 
 We refer to~\cite{maude-book} for more details  on the syntax of
  Real-Time Maude. 

A \emph{class} declaration
\quad
  @class@ \(C\) @|@ \(\attone\) @:@ \(\sonex\)@,@ \dots @,@ \(\attn\) @:@ \(\sn\) 
\quad 
declares a class $C$ with attributes $att_1$ to $att_n$ of
sorts $s_1$ 
to $s_n$. An {\em object\/} of class $C$  is 
represented as a term
$@<@\: O : C \mid att_1: val_1, ... , att_n: val_n\:@>@$
 where $O$  is the
object's
\emph{identifier}, and where $val_1$ to 
$val_n$ are the current values of the attributes $att_1$ to
$att_n$.
The global 
 state has the form @{@$t$@}@, where $t$ 
 is a term of 
 sort @Configuration@ that     has 
the structure of a  \emph{multiset}  of objects and messages, with 
multiset union  denoted by a juxtaposition
operator  that is declared associative and commutative.
%
A \emph{subclass} inherits all the attributes and rules of its 
superclasses.

The dynamic behavior of concurrent
object systems is axiomatized by specifying each of its transition patterns by a rewrite rule. 
For example, the rule

\small
 \begin{alltt}
 rl [l] : m(O,w)  < O : C | a1 : x, a2 : O', a3 : z >   =>
                  < O : C | a1 : x + w, a2 : O', a3 : z >  dly(m'(O'),x) .
\end{alltt}
\normalsize

 \noindent  defines a parametrized family of transitions 
  in which a 
 message @m@, with parameters @O@ and @w@, is read and
 consumed by an object @O@ of class @C@. The transitions change 
 the attribute @a1@ of the  object @O@ and  send a new message
 @m'(O')@ \emph{with delay} @x@.  
 ``Irrelevant'' attributes (such as @a3@)
 need not be mentioned in a rule.

A 
Real-Time Maude specification is \emph{executable}, and the tool 
provides a variety of formal analysis methods. 
The \emph{rewrite} command 
\quad@(trew @$t$@ in time <= @$\tau$@ .)@\quad 
simulates \emph{one} 
behavior of the system within time $\tau$, starting with a given initial state  $t$.
The \emph{search} command 

\begin{alltt}
(tsearch [\(n\)] \(t\) =>* \(pattern\) such that \(cond\) in time <= \(\tau\) .)
\end{alltt}

\noindent
uses a breadth-first strategy to find $n$ states reachable from the initial state $t$ within time $\tau$,
which match a \emph{pattern} and satisfy a \emph{condition}.
The Real-Time Maude's \emph{LTL model checker} 
   checks whether
each behavior from an initial state, possibly  up to a  time bound,
  satisfies a linear temporal logic 
  formula.
 \emph{State propositions}  are 
operators of sort @Prop@. 
%
%
%
A temporal logic \emph{formula} is constructed by state
propositions and
temporal logic operators such as @True@, @~@ (negation),
@/\@, @\/@, @->@ (implication), @[]@ (``always''), @<>@
(``eventually''), @U@ (``until''), and @O@ (``next''). 
The command 
\quad
 @(mc@ \(\,t\) @|=u@ \(\,\varphi\)@ in time <= @\(\tau\) @.)@
\quad 
  checks whether
the temporal logic formula $\varphi$ holds in all behaviors up to
duration $\tau$
 from the initial state $t$.

\section{The Airplane Turning Control System}
\label{sec:airplane-control}

\begin{figure}[ht]
\centering
\includegraphics[width=0.5\textwidth] {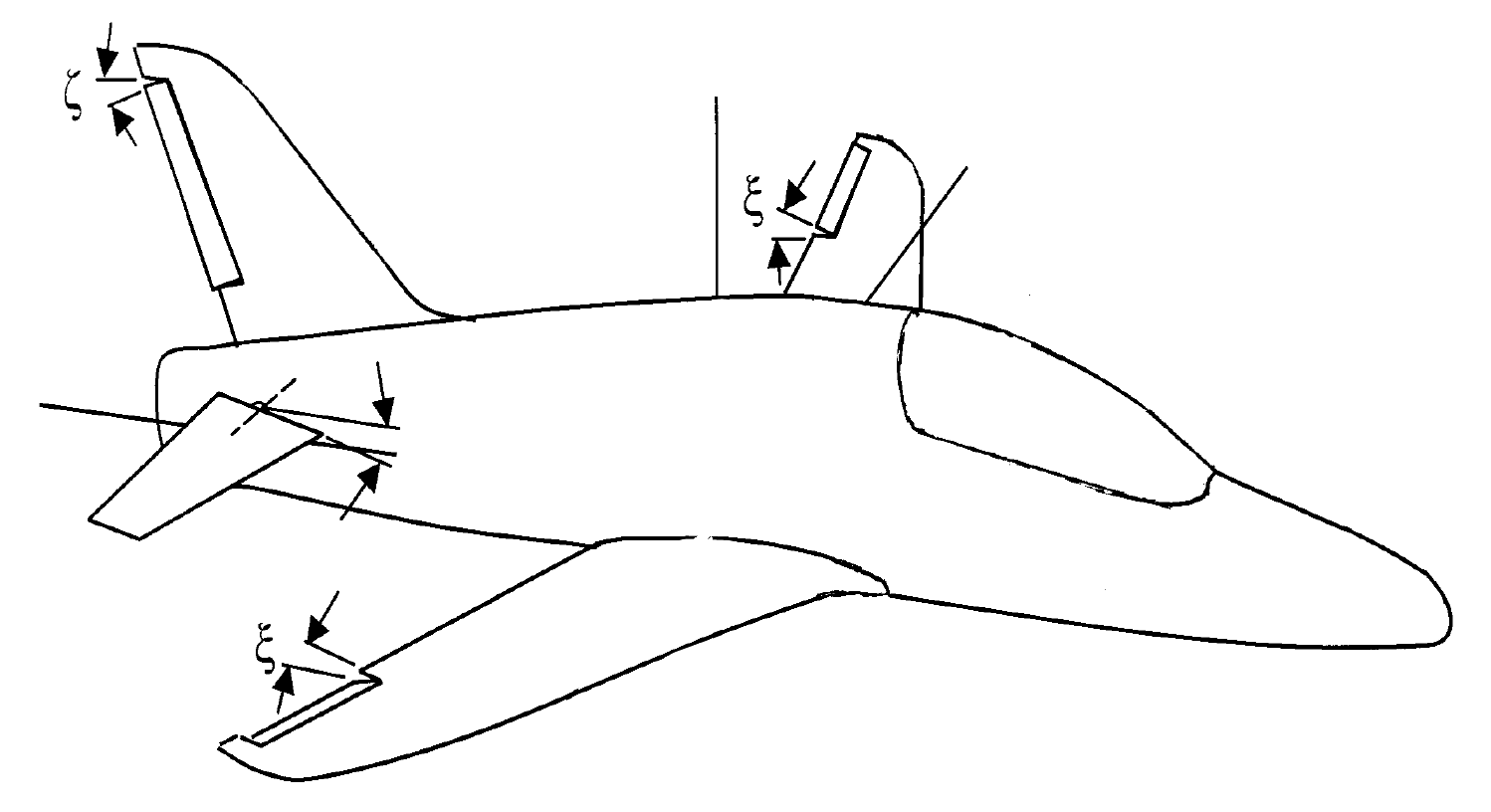}    
\caption{The ailerons and the rudder of an aircraft.}  \label{fig:airplane-str}
\end{figure}

This section presents a simple model of an avionics control system to turn an aircraft.
In general,
the direction of an aircraft is maneuvered by 
the ailerons and the rudder.
As shown in Figure~\ref{fig:airplane-str}, an aileron is a flap attached to the end of the left or the right wing,
and a rudder is a flap attached to the vertical tail 
(the aircraft figures in this section are borrowed from~\cite{collinson1996introduction}).

\begin{figure}[ht]
\centering
\includegraphics[trim=0.3cm 0.4cm 0.3cm 0.4cm, clip=true, width=\textwidth]{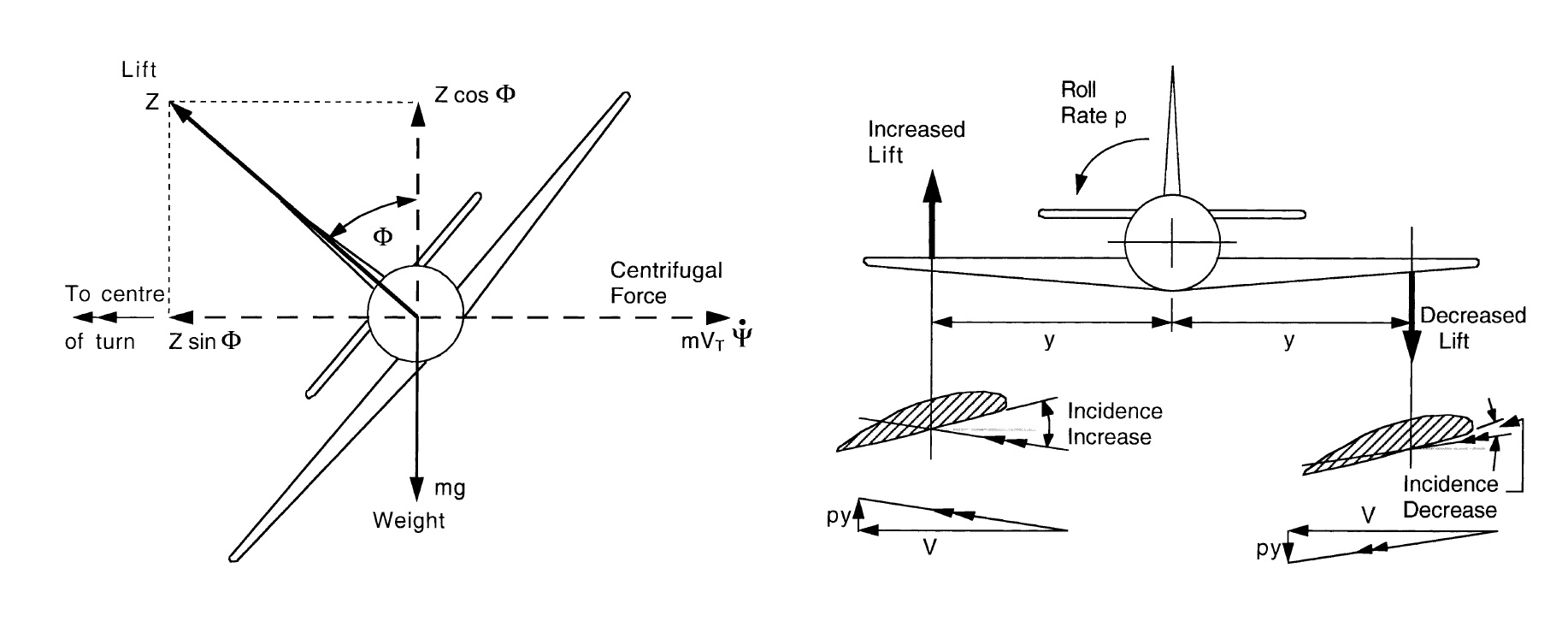}    
\caption{Forces acting in a turn of an aircraft with $\phi$ a roll angle and $\rho$ a roll rate.}  \label{fig:turn}
\end{figure}

When an aircraft makes a turn, the aircraft rolls towards the desired direction of the turn,
so that the lift force caused by the two wings acts as the centripetal force and the aircraft moves in a circular motion.
The turning rate $d\psi$  can be given as a function of the aircraft's roll angle $\phi$:
\begin{equation}\label{eqn:dir}
	d\psi = (g / v) \,*\, \tan \phi
\end{equation}
where $\psi$ is the direction of the aircraft,  $g$ is the gravity constant, and $v$ is the velocity of the aircraft \cite{collinson1996introduction}.
The ailerons
are used to control the rolling angle $\phi$ of the aircraft by
generating different amounts of lift force in the left and the right
wings. Figure~\ref{fig:turn} describes such a banked turn using the
ailerons.

\begin{wrapfigure}{r}{0.34\textwidth}
\centering
\includegraphics[trim=0.2cm 0.1cm 0.2cm 0.2cm, clip=true,width=0.30\textwidth]{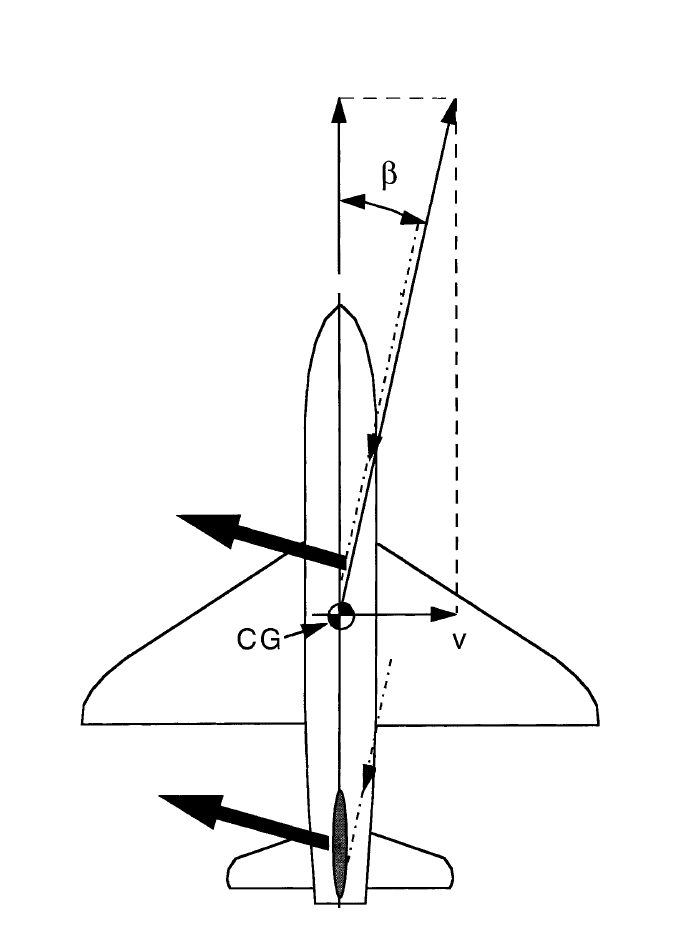}    
\caption{Adverse yaw. \label{fig:yaw}}
\end{wrapfigure}

However, the rolling of the aircraft causes a difference in drag on
the left and the right wings, which produces a yawing moment in the
opposite direction to the roll, 
called adverse yaw.
This adverse yaw makes the aircraft sideslip in a wrong direction 
with the amount of the yaw angle $\beta$, as described in Figure~\ref{fig:yaw}.
This undesirable side effect is usually countered by using the aircraft's rudder,
which generates the side lift force on the vertical tail that opposes the adverse yaw.
To turn an aircraft safely and effectively, the roll angle $\phi$ of the aircraft should be increased for the desired direction
while the yaw angle $\beta$ stays at $0$.

Such a roll and yaw can be modeled by simple mathematical equations
under some simplifying assumptions, including:
\begin{inparaenum}[(i)]
	\item the aircraft's wings are flat with respect to the horizontal axis of the aircraft,
	\item the altitude of the aircraft does not change, which can
          be separately controlled by using the aircraft's elevator (a~flap attached to the horizontal tail of the aircraft),
	\item the aircraft maintains a constant speed by separately controlling the thrust power of the aircraft,
	 and
	\item there are no external influences such as  wind or  turbulence.
\end{inparaenum}
Then, the roll angle $\phi$ and the yaw angle $\beta$ can be modeled by the following equations~\cite{anderson2005introduction}:

\begin{align}
	d\phi^2 &= (\mathit{Lift\ Right} \,-\, \mathit{Lift\ Left}) \,/\, (\mathit{Weight} \,*\, \mathit{Length\ of\ Wing}) \label{eqn:roll}\\
	d\beta^2 &= \mathit{Drag\ Ratio} \,*\, (\mathit{Lift\ Right} \,-\, \mathit{Lift\ Left}) \,/\, (\mathit{Weight} \,*\, \mathit{Length\ of\ Wing}) \nonumber\\
	&\ + \mathit{Lift\ Vertical} \,/\, (\mathit{Weight} \,*\, \mathit{Length\ of\ Aircraft})  \label{eqn:yaw}
\end{align}
The lift force from the left, the right, or the vertical wing is given by the following linear equation:
\begin{equation}
	\mathit{Lift} = \mathit{Lift\ constant} \;*\; \mathit{Angle} 
\end{equation}
where, for $\mathit{Lift\ Right}$ and $\mathit{Lift\ Left}$, $\mathit{Angle}$ is the angle of the aileron,
and for $\mathit{Lift\ Vertical}$,  $\mathit{Angle}$ is the angle of the rudder.
The lift constant depends on the geometry of the corresponding wing,
and the drag ratio is given by the size and the shape of the entire aircraft.
%
%

\begin{figure}[htb]
\begin{center}
\includegraphics[width=0.9\textwidth]{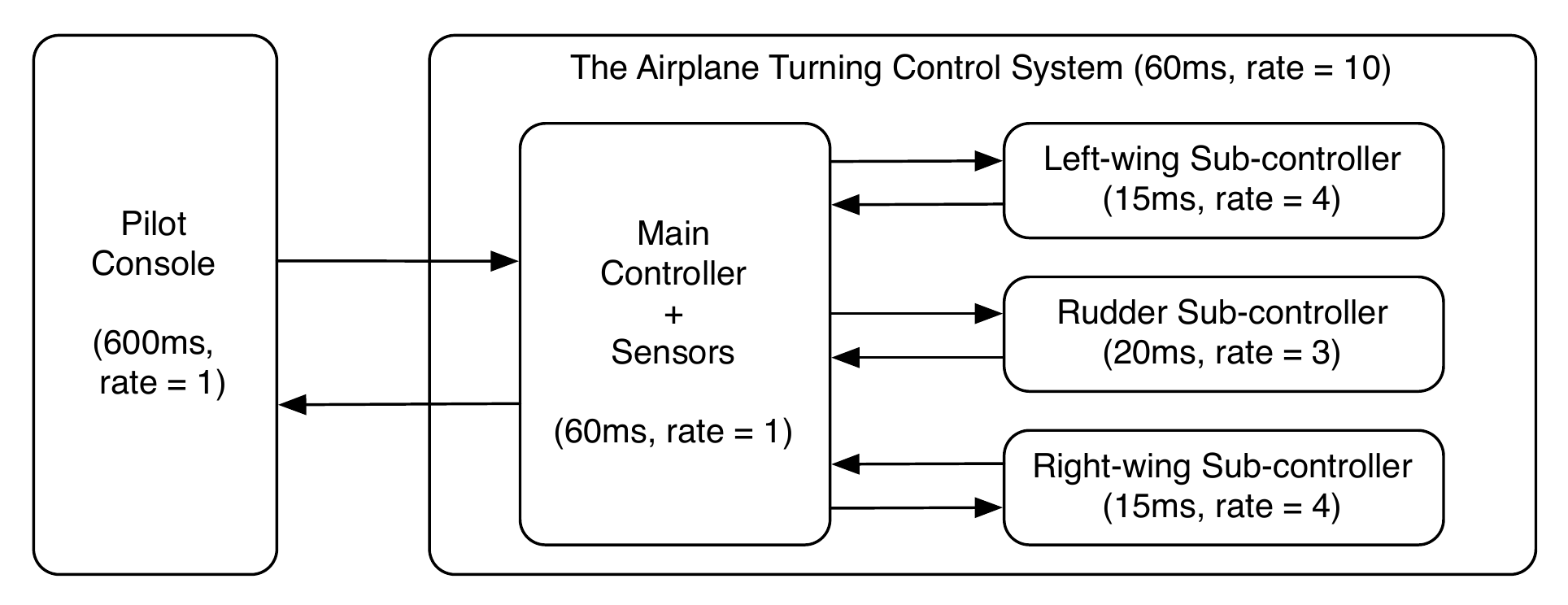}
\caption{The architecture of our airplane turning control system.\label{fig:airplane}}
\end{center}
\end{figure}

We model the airplane turning control system as a multirate ensemble 
with $4$ typed machines: the main controller, the left wing controller, 
the right wing controller, and the rudder controller.
The environment for the airplane turning control system is given by the pilot console,
which is  modeled as another typed machine
and is connected to the main controller on the outside of the control system.
 Each sub-controller moves the surface of the wing towards the goal angle specified 
 by the main controller, which sends the desired angles to the sub-controllers to make 
 a coordinated turn whose goal direction is specified  from the pilot console.
 The main controller also models position sensors that measure the roll, the yaw, and the direction
 by using the aeronautics equations above. 
  In this case study,
 we assume that the main controller has period $60\,\mathrm{ms}$, 
the left and the right wing controllers have period $15\,\mathrm{ms}$ (rate~$4$),
and the rudder controller has period $20\,\mathrm{ms}$ (rate~$3$).
Our model  is a two-level hierarchical multirate ensemble,
since the airplane turning control system itself forms a single typed machine with period $60\,\mathrm{ms}$ (rate~$10$)
 and then is connected to the pilot console with period $600\,\mathrm{ms}$,
 as illustrated in Figure~\ref{fig:airplane}.

Using the framework introduced in Section 4 for specifying and
executing multirate synchronous ensembles in Real-Time Maude, 
we  specify in Section 5 (and redefine in Section 6)  the multirate ensemble $\mathfrak{E}$ corresponding to 
the above airplane control system.
In Section 6 we then  exploit the bisimulation $\mathfrak{E} \,\simeq\, \MA(\mathfrak{E},T,\Gamma)$
(see \cite{bae-facs12}) to verify properties
about the asynchronous realization $\MA(\mathfrak{E},T,\Gamma)$ by model checking them on the much simpler system $\mathfrak{E}$.

\section{Multirate Synchronous Ensembles in Real-Time Maude}
\label{sec:framework}

We have defined a framework for formally modeling and analyzing
multirate ensembles in Real-Time Maude. Given a specification of
\emph{deterministic} 
single typed machines, their periods, input adaptors, a wiring
diagram, and one top-level nondeterministic environment,
our framework gives an executable Real-Time Maude model
of the synchronous composition of the ensemble. 
It is natural to assume that controller components are  deterministic;
however, if we have a nondeterministic component, our framework could
be modified 
so that local transitions are modeled by rewrite rules instead of equationally defined functions.
If the system has  ``local'' fast environments,
they should be dealt with by the corresponding fast machines.%
\footnote{An environment can  be viewed
  as a nondeterministic typed machine~\cite{pals-tcs}. Therefore, a
  faster machine's 
  environment  and the fast machine  itself form a nondeterministic
  2-machine ensemble.}  

This section gives a
brief overview of our framework and of how the user should specify
the ensemble; the entire definition of our framework is given in our
longer report~\cite{ftscs12-techrep}. 

\paragraph{Representing Multirate Ensembles in Real-Time Maude.}

The multirate  ensemble can be naturally specified in an object-oriented style,
where its machines and the ensemble can be modeled as objects.
A~typed machine is represented as an object instance of a subclass
of the base class \texttt{Component},
which has the common attributes @rate@, @period@, and @ports@.

\small
\begin{alltt}
class Component | rate : NzNat,   period : Time,   ports : Configuration .
\end{alltt}
\normalsize

\noindent
The @period@ denotes the period of  the typed machine, and 
 @rate@  denotes its rate in a multirate ensemble.
The \texttt{ports} attribute contains the input/output "ports"  of a typed machine,
represented as a multiset of \texttt{Port} objects, whose
\texttt{content} attribute contains the data content as 
a list of values of the supersort  \texttt{Data}:

\small
\begin{alltt}
class Port | content : List\char123Data\char125 .   class InPort .   class OutPort .
subclass InPort OutPort < Port .
\end{alltt}
\normalsize

For each typed machine, the user must define an appropriate subclass
of the class \texttt{Component}, 
the function \texttt{delta} defining its transition function, 
and the \texttt{adaptor} function for each input port:

\small
\begin{alltt}
op delta : Object -> Object .
op adaptor : ComponentId PortId NeList\char123Data\char125 -> NeList\char123Data\char125 .
\end{alltt}
\normalsize

\noindent
where the sort @NeList{Data}@ denotes a non-empty list of data.
%
%
%
%
%
%
%

A multirate machine ensemble is modeled 
as an object instance of the class \texttt{Ensemble}.
We  support the definition of hierarchical ensembles by letting an
ensemble be a @Component@, 
which also contains the wiring diagram (\texttt{connections})
and the \texttt{machines} in the ensemble.
In this case, the \texttt{ports} attribute represents
the environment ports of the ensemble:

\small
\begin{alltt}
class Ensemble | machines : Configuration,   connections : Set\char123Connection\char125 .
subclass Ensemble < Component .
\end{alltt}
\normalsize 

A wiring diagram  is modeled as a set of connections.
A  connection from an output port $P_1$ of a component $C_1$ to an
input port $P_2$ 
of a component $C_2$ is represented as a term
\texttt{\(C\sb{1}\).\(P\sb{1}\) --> \(C\sb{2}\).\(P\sb{2}\)}. 
Similarly, 
a connection between an environment port $P_1$ and a machine port $P_2$
of a subcomponent $C_2$ is represented as a term \texttt{\(P\) -->
  \(C\sb{2}\).\(P\sb{2}\)} (for an environment input) 
or a term \texttt{\(C\sb{2}\).\(P\sb{2}\) --> \(P\)} (for an environment output).

%

\paragraph{Defining Synchronous Compositions of Multirate Ensembles.}

Given these definitions of an ensemble, our framework defines its synchronous
composition as follows. 
The function \texttt{delta} of a multi-rate machine
ensemble~$\mathfrak{E}$  defines the transitions in the 
synchronous composition of $\mathfrak{E}$.

\small
\begin{alltt}
eq delta(< C : Ensemble | >)
 = transferResults(execute(transferInputs(< C : Ensemble | >))) .
\end{alltt}
\normalsize

\noindent
In the above equation, the function \texttt{transferInputs}  first
transfers to each input port 
a value in the corresponding environment output port or the feedback output port.
The \texttt{execute} function below  applies  the appropriate input adaptor to each sub-component,
and then performs the function \texttt{delta} of each component as
many times as its deceleration rate.
Finally, the new outputs in sub-components are transferred to 
the environment ports by the function \texttt{transferResults}.

\small
\begin{alltt}
eq execute(< C : Ensemble | machines : COMPS >)
 = < C : Ensemble | machines : executeSub(COMPS) > . 
eq executeSub(< C : Component | > COMPS)
 = k-delta(applyAdaptors(< C : Component | >)) executeSub(COMPS) .
eq executeSub(none) = none .
\end{alltt}
\normalsize

\noindent
where 
 \texttt{k-delta}  applies \texttt{delta} 
as many times as the rate of a given typed machine:

\small
\begin{alltt}
eq k-delta(< C : Component | rate : N >) = k-delta(N, < C : Component | >) .
eq k-delta(s N, OBJECT) = k-delta(N, delta(OBJECT)) .
eq k-delta(  0, OBJECT) = OBJECT .
\end{alltt}
\normalsize

\noindent
The rates and periods  should be consistent in an ensemble; 
that is, if some component has rate $k$ and period $p$, 
then any component with rate $k \cdot t$ in the same ensemble should have period $p / t$.

\paragraph{Formalizing the Synchronous Steps.}

When each typed machine is deterministic
and the system contains one (nondeterministic) top-level environment,
the dynamics of the entire system given by a multirate machine
ensemble is specified  
by the following  tick rewrite rule
that simulates each synchronous step of the composed system:

\small
\begin{alltt}
crl [step] : \char123< C\! : Component | period : T >\char125 
           => 
             \char123delta(envOutput(ENVOUTPUT, clearOutputs(< C\! : Component | >)))\char125 
            in time T
 if possibleEnvOutput => ENVOUTPUT .
\end{alltt}
\normalsize

In  the condition of the rule, \emph{any}  possible environment output
can be nondeterministically assigned to 
the variable \texttt{ENVOUTPUT}, since the 
constant @possibleEnvOutput@ can be rewritten to any possible
environment output 
by user-defined rewriting rules of the form 

\small
\begin{alltt}
rl possibleEnvOutput => (PortId\(\sb{1}\) = Value\(\sb{1}\), \(\ldots\), PortId\(\sb{n}\) = Value\(\sb{n}\)) .
\end{alltt}
\normalsize

\noindent
These  values are inserted into the appropriate component input port by the \texttt{envOutput} function,
after clearing the outputs generated in the previous round
by the @clearOutputs@ function. 
 
\footnotesize
\begin{alltt}
eq envOutput((P = DL,\;ENVASSIGNS),  < C\! : \!Component | ports\! : \!< P\! : \!InPort | > PORTS >)
 = envOutput(ENVASSIGNS,  < C\! : \!Component | ports\! : \!< P\! : \!InPort | content\! : \!DL > PORTS >) .
eq envOutput(empty, < C\! : \!Component | >) = < C\! : \!Component | > .
\end{alltt}
\normalsize

The \texttt{delta} function is finally applied to perform the transition of the component.
Since an ensemble is an instance of the @Component@ class
and the @delta@ function is also given for the @Ensemble@ class,
this  tick rewrite rule can also be applied to a hierarchical multirate ensemble.
After each synchronous step, 
the input ports of the component contain the environment input given by 
the @possibleEnvOutput@,
and the output ports will have the resulting environment output generated by 
the @delta@ function.
The @period@ does not have any effect on the structure of
the transition system, 
but  can be useful to verify \emph{timed} properties, such as
time-bounded LTL properties
and metric CTL properties~\cite{lepri-wrla12}.

\section{Modeling the Airplane Turning Control System}
\label{sec:airplane-model}

In this section we formally specify the airplane turning control
system in Section~\ref{sec:airplane-control}
using the ensemble framework in Real-Time Maude described in
Section~\ref{sec:framework}. 
The entire specification is available in our report~\cite{ftscs12-techrep}.
%
%
The following parameters are
chosen to be representative of a small general aviation aircraft.
The speed of the aircraft is assumed to be $50\, m /s$, 
and the gravity constants is $g = 9.80555\, m / s^2$.

\small
\begin{alltt}
eq planeSize     = 4.0 .    eq weight        = 1000.0 .     eq wingSize  = 2.0 .
eq virtLiftConst = 0.6 .    eq horzLiftConst = 0.4 .        eq dragRatio = 0.05 .       
\end{alltt}
\normalsize

\paragraph{Subcontroller.}

The subcontrollers for the ailerons and the rudder are
modeled as object instances of the following class
\texttt{SubController}:

\small
\begin{alltt}
class SubController | curr-angle : Float,   goal-angle : Float,   diff-angle : Float .
subclass SubController < Component .  
\end{alltt}
\normalsize  

A subcontroller increases/decreases the @curr-angle@ toward the @goal-angle@
in each round, but the difference in a single (fast) round should be
less than or equal to 
the maximal angle @diff-angle@.
The transition function @delta@ of a subcontroller is then defined by
the following equation: 

\footnotesize
\begin{alltt}
ceq delta(< C : SubController | ports : < input : InPort | content : D LI >
                                        < output : OutPort | content : LO >,
                                curr-angle : CA, goal-angle : GA, diff-angle : DA >)
  = 
      < C : SubController | ports : < input : InPort | content : LI > 
                                    < output : OutPort | content : LO d(CA') >,
                            curr-angle : CA', goal-angle : GA' > 
  if CA' := angle(moveAngle(CA,GA,DA))
  /\char92 GA' := angle(if D == bot then GA else float(D) fi) .
\end{alltt}
\normalsize

\noindent where  \texttt{moveAngle(CA,\(\,\)GA,\(\,\)DA)}
equals  the angle that is increased or decreased from 
the current angle \texttt{CA} to the goal angle \texttt{GA} 
up to the maximum angle difference \texttt{DA}:

\small
\begin{alltt}
eq moveAngle(CA,\,GA,\,DA) = if abs(GA\;-\;CA) > DA then CA + DA * sign(GA\;-\;CA) else GA fi .
\end{alltt}
\normalsize
   
\noindent
The \texttt{angle} function keeps the angle value between $-180^\circ$
and $180^\circ$.  
The \texttt{delta} function updates the goal angle
to the input from the main controller, 
and  keeps the previous goal if it receives @bot@ (i.e., $\bot$).

\paragraph{Main Controller.}

The main controller for the aircraft turning control system
 is modeled as an  object instance of the following class
 \texttt{MainController}:

\small
\begin{alltt}
class MainController | velocity : Float,   goal-dir : Float,
                       curr-yaw : Float,   curr-rol : Float,    curr-dir : Float .
subclass MainController < Component .
\end{alltt}
\normalsize  

\noindent
The \texttt{velocity} attribute denotes  the speed of the aircraft. 
The \texttt{curr-yaw}, \texttt{curr-roll}, and \texttt{curr-dir}
attributes model the position sensors of the aircraft by indicating 
 the current  yaw angle $\beta$, 
roll angle $\phi$, and  direction $\Psi$, 
respectively.
The \texttt{goal-dir} attribute denotes
 the goal direction given by the pilot.

For each round of the main controller,
the  attributes \texttt{curr-yaw}, \texttt{curr-roll}, and \texttt{curr-dir}
are updated%
\footnote{
We currently use the simple Euler's method to compute such position values
given by the differential aeronautical equations  (\ref{eqn:dir}-\ref{eqn:yaw}),
but more precise methods can be easily applied.}
 using the angles of the wings in the input ports
that are transferred from the subcontrollers.
The \texttt{goal-dir} is also updated if a new goal direction arrives
to the input ports. 
Based on the new current position status and the goal direction,
the new angles of the wings are evaluated and sent back to the subcontrollers. 
The transition function @delta@ of the main controller is then defined as follows:

\footnotesize
\begin{alltt}
ceq delta(< C\! : \!MainController | 
              ports : < input\! : \!InPort | content\! : \!IN PI >    < output\! : \!OutPort | content\! : \!PO >
                      < inLW\! : \!InPort | content\! : \!d(LA) LI >   < outLW\! : \!OutPort | content\! : \!LO >
                      < inRW\! : \!InPort | content\! : \!d(RA) RI >   < outRW\! : \!OutPort | content\! : \!RO >
                      < inTW\! : \!InPort | content\! : \!d(TA) TI >   < outTW\! : \!OutPort | content\! : \!TO >,
              velocity\! : \!VEL, period\! : \!T, 
              curr-yaw\! : \!CY, curr-rol\! : \!CR, 
              curr-dir\! : \!CD, goal-dir\! : \!GD >)
  =    
      < C\! : \!MainController | 
          ports : < input\! : \!InPort | content\! : \!PI >  < output\! : \!OutPort | content\! : \!PO OUT >
                  < inLW\! : \!InPort | content\! : \!LI >   < outLW\! : \!OutPort | content\! : \!LO d(-\;RA') >
                  < inRW\! : \!InPort | content\! : \!RI >   < outRW\! : \!OutPort | content\! : \!RO d(RA') >
                  < inTW\! : \!InPort | content\! : \!TI >   < outTW\! : \!OutPort | content\! : \!TO d(TA') >,
          curr-yaw\! : \!CY', curr-rol\! : \!CR', 
          curr-dir\! : \!CD', goal-dir\! : \!GD' > 
   if CY' := angle( CY + dBeta(LA,RA,TA) * float(T) )
   /\char92 CR' := angle( CR + dPhi(LA,RA) * float(T) )
   /\char92 CD' := angle( CD + dPsi(CR,VEL) * float(T) )
   /\char92 GD' := angle( if IN == bot then GD else GD + float(IN) fi )
   /\char92 RA' := angle( horizWingAngle(CR', goalRollAngle(CR', CD', GD')) )
   /\char92 TA' := angle( tailWingAngle(CY') )
   /\char92 OUT := dir: CD' roll: CR' yaw: CY' goal: GD' .
\end{alltt}
\normalsize

\noindent
The first four lines in the condition compute new values for
\texttt{curr-yaw}, \texttt{curr-roll}, \texttt{curr-dir}, 
and \texttt{goal-dir}, based on values in the input ports. 
A non-$\bot$ value in the port @input@ is added to @goal-dir@.
The variables @RA'@ and @TA'@ denote new angles of the ailerons and the rudder,
computed by the control functions explained below.
Such new angles are queued in the corresponding output ports,
and will be transferred to the related subcontrollers at the next synchronous step
since they are feedback outputs.
The last line in the condition gives the output for the current step,
the new position information of the aircraft,
which will be transferred to its container ensemble at the end of the current synchronous step.

The new angles of the ailerons and the rudder are computed by the
\emph{control functions}. 
The function \texttt{horizWingAngle}  computes the new angle for the
aileron in the right wing, 
based on the current roll angle and the goal roll angle. 
The angle of the aileron in the left wing is always exactly opposite
to the one of the right wing. 
The function \texttt{goalRollAngle}  computes the desired roll angle $\phi$
to make a turn,
based on the current roll angle and the difference between the goal
direction and the current direction. 
Finally, in order to achieve a coordinated turn where the yaw angle is always $0$,
the function \texttt{tailWingAngle} computes
the new rudder angle based on the current yaw angle. 
We define all three control functions by simple linear equations as follows,
where @CR@ is a current roll angle and @CY@ is a current yaw angle:

\small
\begin{alltt}
eq goalRollAngle(CR,CD,GD) = sign(angle(GD\,-\,CD)) * min(abs(angle(GD\,-\,CD))\;*\;0.3, 20.0) .
eq horizWingAngle(CR,GR)   = sign(angle(GR\,-\,CR)) * min(abs(angle(GR\,-\,CR))\;*\;0.3, 45.0) .
eq tailWingAngle(CY)       = sign(angle(- CY)) * min(abs(angle(- CY))\;*\;0.8, 30.0) .
\end{alltt}
\normalsize

\noindent
That is, the goal roll angle is proportional to the difference
$\texttt{GD}\,-\,\texttt{CD}$ between the goal and current directions 
with the maximum $20^\circ$.
The horizontal wing (aileron) angles are also proportional to the
difference $\texttt{GR}\,-\,\texttt{CR}$ between the goal and current
roll angles 
with the maximum $45^\circ$.
Similarly, the rudder angle is proportional to the difference
$-\,\texttt{CY}$ between the goal and current yaw angles 
with the maximum $30^\circ$, where the goal yaw angle is always $0^\circ$.

\paragraph{Pilot Console.}

The pilot console,
the environment for the aircraft turning control system,
 is modeled as an  object instance of the following class \texttt{PilotConsole}:

\small
\begin{alltt}
class PilotConsole | scenario : List\char123Data\char125 .
subclass PilotConsole < Component .
\end{alltt}
\normalsize    

The attribute @scenario@ contains a list of goal angles that are
transmitted to the main controller. 
The transition function @delta@ of the pilot console
keeps sending goal angles in the @scenario@ to its output port
until no more data remains in the @scenario@.
In addition, the pilot console has an extra input port to receive
an outer environment input to generate \emph{nondeterministic} goal directions.
A non-$\bot$ value in the input port is added to
the output given by the @scenario@. 

\footnotesize
\begin{alltt}
ceq delta(< C : PilotConsole | ports : < input : InPort | content : IN LI >
                                       < output : OutPort | content : LO >,
                               scenario : d(F) SL >)
  = 
    < C : PilotConsole | ports : < input : InPort | content : LI >
                                 < output : OutPort | content : LO OUT >,
                         scenario : SL > .
 if OUT := d(if IN == bot then F else angle(F + float(IN)) fi) .

eq delta(< C : PilotConsole | ports : < input : InPort | content : IN LI >
                                      < output : OutPort | content : LO >,
                              scenario : nil >)
 = 
   < C : PilotConsole | ports : < input : InPort | content : LI >
                                < output : OutPort | content : LO bot >  > .
\end{alltt}
\normalsize 

\noindent
If the @scenario@ is empty, i.e., @nil@, the outer
environment input is ignored, 
and the pilot console will keep sending $\bot$ to its output port.

\paragraph{Airplane System.}

The entire architecture of the airplane turning control system in
Figure~\ref{fig:airplane}, 
 including the environment (i.e., the pilot console),
is then represented as an ensemble 
object with subcomponents as follows (some parts of the specification
are replaced by `\texttt{\ldots}'): 

\footnotesize
\begin{alltt}
< airplane : \emph{Ensemble} | 
   rate : 1, 
   period : 600,
   ports :  < input : InPort | content : nil >  < output : OutPort | content : nil >,
   connections :
      input --> pilot\,.\,input ;  pilot\,.\,output --> csystem\,.\,input ;  csystem\,.\,output --> output,
   machines :
      \(\big(\)< pilot : \emph{PilotConsole} | rate : 1, period : 600,\ldots >
       < csystem : \emph{Ensemble} |
          rate :10, period : 60,
          ports : < input : InPort | content : nil >  
                  < output : OutPort | content : nil >,
          connections :
             input --> main\,.\,input ;  main\,.\,output --> output ; 
             left\,.\,output --> main\,.\,inLW ;  main\,.\,outLW --> left\,.\,input ; \ldots,
          machines : (< main : \emph{MainController} | rate : 1, period : 60,\ldots >
                      < left : \emph{SubController} | rate : 4, period : 15,\ldots >
                      < right : \emph{SubController} | rate : 4, period : 15,\ldots >
                      < rudder : \emph{SubController} | rate : 3, period : 20,\ldots >) >\(\big)\) >
\end{alltt}
\normalsize   

\noindent
The top-level  ensemble @airplane@ includes 
the pilot console @pilot@ and the ensemble @csystem@ for the
airplane turning control system.
The @airplane@ has one input port for the pilot console to generate
nondeterministic goals, 
and one output port to display the result.

In the ensemble @csystem@,
the input adaptors for the subcontrollers generate a~vector with extra $\bot$'s
and the adaptor for the main controller selects the last value of the input vector.

\small
\begin{alltt}
eq adaptor(left, input, D) = D bots(3) .   eq adaptor(rudder, input, D) = D bots(2) .
eq adaptor(right, input, D) = D bots(3) .  eq adaptor(main, P, LI D) = D .
\end{alltt}
\normalsize

\noindent
The function @bots(n)@ generates $n$ $\bot$-constants.
Similarly, in the top ensemble @airplane@, 
the input adaptor for the ensemble @csystem@
generates a vector with extra $\bot$'s,
and the adaptor for the pilot console selects the last value of the input vector.

\small
\begin{alltt}
eq adaptor(csystem, input, D) = D bots(9) 
eq adaptor(pilot, input, LI D) = D .
\end{alltt}
\normalsize

\section{Formal Analysis of the Airplane Turning  Control System }
\label{sec:analysis}

This section explains  
how we have formally analyzed the above Real-Time
Maude model of the multirate synchronous design  of the airplane
turning control system, 
and how the turning control system  has been improved as a result of our analysis.
Multirate PALS  ensures that the verified properties
also hold in a distributed real-time realization of the design.
There are two important requirements that 
the airplane turning control system  should satisfy:
\begin{itemize}
	\item \emph{Liveness}: 
	the airplane should reach the goal direction within a reasonable time with a stable status 
	in which both the roll angle and the yaw angle are close to $0$.за
	\item \emph{Safety}: during a turn, the yaw angle should always be close to $0$.
\end{itemize}

We first analyze deterministic behaviors when the airplane turns $+ 60^\circ$ to the right.%
\footnote{In our model, a turn of positive degrees is a right turn, and one of negative degrees a left turn.}
In this case, there are no outer environment inputs to the pilot console,
so that the pilot console sends the goal angles in the scenario to the main controller in turn.
We consider the following three scenarios:

\begin{enumerate}
	\item The pilot gradually increases the goal direction in 6 seconds, $+10^\circ$ for each second.
	\item The pilot sets the goal direction to $+ 60^\circ$ at the first step.
	\item\label{item:scenario-3}
	 The goal direction is at first $- 30^\circ$, and then it is suddenly set to $+ 60^\circ$ in one second.
\end{enumerate}

\begin{figure}[htb]
\centering
\includegraphics[trim=0.9cm 0.7cm 0.7cm 0.9cm, width=\textwidth] {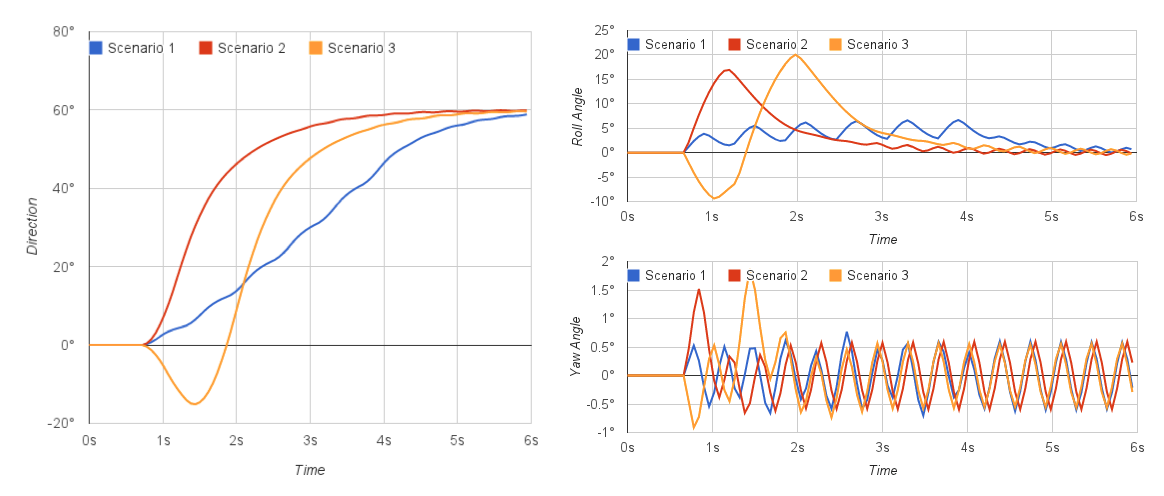}    
\caption{Simulation results for turning scenarios: the directions of
  aircraft (left), the roll angles (top right), and the yaw angles
  (bottom right)}  \label{fig:simul} 
\end{figure}

\noindent
Figure~\ref{fig:simul} shows the simulation results for these three scenarios up to $6$ seconds,
obtained by using the Real-Time Maude simulation command 
\quad\texttt{(trew
  \char123 model(\(\mathit{Scenario}\)){\char125} in time <= 6000
  .)}, 
where  \texttt{model(\(\mathit{Scenario}\))} gives the initial state of the system
with the scenario $\mathit{Scenario}$ for the pilot console.
For example, the scenario \ref{item:scenario-3} is represented as the term ``@d(-30.0) d(90.0)@'' for a list of directions.
As we can see in the graph, the airplane reaches the goal direction $60^\circ$ in a fairly short time,
and the roll angle also goes to a stable status.
However, the yaw angle seems to be quite unstable.

There are basically two reasons why the yaw angle is not sufficiently close to $0$ during a turn.
First, since all control functions are linear, the new angles for the ailerons and the rudder are not small enough 
when the yaw angle is near $0$. Second, the roll angle is sometimes changing too fast, so that the rudder
cannot effectively counter the adverse yaw.
Therefore, we modify the control functions as follows:

\small
\begin{alltt}
ceq horizWingAngle(CR, GR) 
  = sign(FR) * (if abs(FR)\;>\;1.0 then min(abs(FR)\;*\;0.3, 45.0) else FR\;\emph{^\;2.0}\;*\;0.3 fi) 
 if FR := angle(GR - CR) . \vspace{1ex}
ceq tailWingAngle(CY) 
  = sign(FY) * (if abs(FY)\;>\;1.0 then min(abs(FY)\;*\;0.8, 30.0) else FY\;\emph{^\;2.0}\;*\;0.8 fi)
 if FY := angle(-\,CY) . \vspace{1ex}
ceq goalRollAngle(CR, CD, GD)
  = if abs(FD\;*\;0.32\;-\;CR) > 1.5 then CR + sign(FD\;*\;0.32\;-\;CR) * 1.5 else FD\;*\;0.32 fi 
 if FD := angle(GD - CD) .
\end{alltt}
\normalsize

\noindent
When the difference between the goal and the current angles (i.e., @FR@ or @FY@) is less than
or equal to~$1$,
the functions @horizWingAngle@ and @tailWingAngle@ are now \emph{not} proportional
to the difference, but proportional to the \emph{square} of it.
Furthermore, the goal roll angle can be changed at most $1.5^\circ$ at a time,
so that there is no more abrupt rolling.

In order to check if the new control functions are safe, we define some auxiliary functions.
The function \texttt{$\mathit{PortId}$ ?= $\mathit{Component}$}
returns the content of the corresponding port in the $\mathit{Component}$:

\small
\begin{alltt}
eq P ?= < C : Component | ports : < P : Port | content : DL > PORTS >  =  DL .
\end{alltt}
\normalsize

\noindent
and the function \texttt{safeYawAll($\mathit{OutputDataList}$)} checks whether every output data 
in the given list has a~safe yaw angle, namely, an angle less than $1.0^\circ$.

\small
\begin{alltt}
eq safeYawAll(DO LI) = abs(yaw(DO)) < 1.0 and safeYawAll(LI) .
eq safeYawAll(nil)   = true .
\end{alltt}
\normalsize

\noindent
Then, we can verify, using the Real-Time Maude search command,
that there is no dangerous yaw angle within sufficient time bound;
e.g., for the scenario \ref{item:scenario-3}:

\small
\begin{alltt}
Maude> \emph{(tsearch [1] \char123model(d(-30.0) d(90.0)){\char125} =>* \char123SYSTEM\char125} 
        \emph{such that not safeYawAll(output ?= SYSTEM) in time <= 27000 .)} \vspace{1ex}
No solution
\end{alltt}
\normalsize

\noindent
Although  each state of the transition system captures only the slow steps for the top ensemble 
(i.e., every $600 \mathrm{ms}$),
 @safeYawAll@ also checks all fast steps for the main controller (every $60 \mathrm{ms}$),
since it accesses the \emph{history} of the main controller's status 
in the output port of the top ensemble,
which the main controller sends to the top ensemble for each fast step of it.

Furthermore, we can use Real-Time Maude's LTL model checking to verify
both liveness and safety  at the same time.
The desired property is that the airplane reaches the desired direction with a stable status while keeping the yaw
angle close to $0$, which can formalized as  the  LTL formula
\[
\square (\neg \texttt{stable} \to (\texttt{safeYaw} \;\mathbf{U}\; (\texttt{reach} \wedge \texttt{stable})))
\]
where the atomic propositions \texttt{safeYaw}, \texttt{stable}, and \texttt{reach}
are defined as follows:

\small
\begin{alltt}
 eq \char123SYSTEM\char125 |= safeYaw = safeYawAll(output ?= SYSTEM) .
 eq \char123SYSTEM\char125 |= stable  = stableAll(output ?= SYSTEM) .
ceq \char123SYSTEM\char125 |= reach   = abs(angle(goal(DO) - dir(DO))) < 0.5 
 if DO := last(output ?= SYSTEM) .
\end{alltt}
\normalsize

\noindent
and the function \texttt{stableAll($\mathit{OutputDataList}$)} returns @true@
only if both the yaw angle and the roll angle are less than $0.5^\circ$ for every output data
in the $\mathit{OutputDataList}$.
We have verified 
that all three scenarios  satisfy the above LTL property with the \emph{new} control functions, 
using the time-bounded LTL model checking command of Real-Time Maude:

\small
\begin{alltt}
Maude> \emph{(mc \char123model(d(-30.0) d(90.0))\char125 |=t [] (~\,stable -> (safeYaw U reach\;/\char92\;stable))} 
           \emph{in time <= 7200 .)}  \vspace{1ex}   
Result Bool : true
\end{alltt}
\normalsize


Finally, we have verified nondeterministic behaviors in which the pilot sends
one of the turning angles $-60.0^\circ$, $-10.0^\circ$, $0^\circ$, $10^\circ$, and $60.0^\circ$
to the main controller for $6$ seconds.
Such nondeterministic behaviors can be defined by adding the following five rewrite rules,
which nondeterministically assign one of these values to the input port of the pilot console:

\small
\begin{alltt}
rl possibleEnvOutput => input = d(0.0) .
rl possibleEnvOutput => input = d(10.0) .   rl possibleEnvOutput => input = d(-10.0) .
rl possibleEnvOutput => input = d(60.0) .   rl possibleEnvOutput => input = d(-60.0) .
\end{alltt}
\normalsize

\noindent
The following model checking command then shows that 
our redesigned system, with the new control functions, satisfies the above LTL property
within $18$ seconds, where one of the above five angles is nondeterministically
chosen and added to the angle $0^\circ$ in the scenario for each step of the pilot console:

\small
\begin{alltt}
Maude> \emph{(mc \char123model(d(0.0) d(0.0) d(0.0) d(0.0) d(0.0) d(0.0))\char125} 
             \emph{|=t} 
           \emph{[] (~ stable -> (safeYaw U (reach\;/\char92\;stable))) in time <= 18000 .)} \vspace{1ex}
Result Bool : true
\end{alltt}
\normalsize

\noindent
The number of states explored in this model checking analysis is $246,785$,%
\footnote{This analysis took almost $9$ hours on Intel Core i5 2.4 GHz with $4$ GB memory.}
which is a huge state space reduction compared to the distributed asynchronous
model since:
\begin{inparaenum}[(i)]
	\item asynchronous behaviors are eliminated thanks to PALS, and
	\item any intermediate fast steps 
		for the sub-components are merged into a single-step of the system's top-level ensemble.
\end{inparaenum}

\section{Conclusions}
\label{sec:concl}

The present work can be seen from different perspectives.  First, from the perspective of research on the PALS methodology, its main contribution is to demonstrate that Multirate PALS, when used in combination with a tool like Real-Time Maude, can be effectively applied to the formal verification of nontrivial DCPS designs, and even to the process of refining a DCPS design before it is verified.  Second, from the perspective of the formal specification and verification of distributed hybrid systems, it also shows that Real-Time Maude is an effective tool for specifying and verifying such systems.

Much work remains ahead.  On the one hand, more case studies like this one should be developed.  On the other hand,
our work in \cite{icfem11}, which
applies  PALS to a synchronous fragment of the AADL CPS modeling language in the single rate case, should be extended to the Multirate PALS case.  Finally, the hybrid system applications of Real-Time Maude should be further developed independently of PALS.  Several such applications have been developed in the past; but many more are possible, and a richer experience will be gained.

\paragraph{Acknowledgments.} 
We thank the anonymous reviewers for their very useful comments that
helped us improve our paper. 
This work was partially supported by Boeing Corporation Grant C8088
and NSF Grant CCF 09-05584.

\bibliographystyle{eptcs}
\bibliography{bibl}


\end{document}